\documentclass[amsmath,amssymb,floatfix,prb,epsf,superscriptaddress,twocolumn]{revtex4}

\usepackage{hyperref}
\usepackage{amsmath,amsfonts,amssymb}
\usepackage{amsthm}
\usepackage{graphicx,color,epsf,epsfig}
\usepackage{lscape}
\usepackage[vcentermath]{youngtab}
\usepackage{rotating}
\usepackage[usenames,dvipsnames]{xcolor}

\def\ds{\displaystyle}
\def\bea{\begin{array}{c}}
\def\ea{\end{array}}
\def\be{\begin{equation}\bea\ds}
\def\ee{\ea\end{equation}}
\def\bee{\begin{equation}\begin{array}{rcl}\ds}
\def\eee{\end{array}\end{equation}}

\def\htau{\hat{\tau}}
\def\hk{\hat{\kappa}}

\def\eb[#1]{{\bf e}_{#1}}
\def\heb[#1]{\hat{\bf e}_{#1}}

\begin{document}

\title{A Continuous Effective Model of the Protein Dynamics}

\author{Dmitry Melnikov}
\affiliation{International Institute of Physics, Federal University of Rio Grande do Norte,  Campus Universit\'ario, Lagoa Nova, Natal-RN  59078-970, Brazil}

\affiliation{Institute for Theoretical and Experimental Physics, B.~Cheremushkinskaya 25, Moscow 117218, Russia}

\author{Alyson B. F. Neves}

\affiliation{Department of Theoretical and Experimental Physics, Federal University of Rio Grande do Norte, Campus Universit\'ario, Lagoa Nova, Natal-RN  59078-970, Brazil}

\affiliation{Federal University of Maranh\~ao, Campus Balsas, rua Jos\'e Le\~ao 484, Balsas-MA 65800-000, Brazil}


\begin{abstract}
The theory of elastic rods can be used to describe certain geometric and topological properties of the DNA molecules. A similar effective field theory approach was previously suggested to describe the conformations and dynamics of proteins. In this letter we report a detailed study of the basic features of a version of the proposed model, which assumes proteins to be very long continuous curves. In the most appealing case, the model is based on a potential with a pair of minima corresponding to helical and strand-like configurations of the curves. It allows to derive several predictions about the geometric features of the molecules, and we show that the predictions are compatible with the phenomenology. While the helices represent the ground state configurations, the abundance of beta strands is controlled by a parameter, which can either completely suppress their presence in a molecule, or make them abundant. The few-parameter model investigated in the letter rather represents a universality class of protein molecules. Generalizations accounting for the discrete nature and inhomogeneity of the molecules presumably allow to model realistic cases.
\end{abstract}


{\onecolumngrid
\hfill{ITEP-TH-nn/yy}
}

\maketitle


Proteins are long quasi one-dimensional chains of amino acids. At the secondary structure level these chains organize themselves in several characteristic configurations, such as helices and beta sheets, appearing due to hydrogen bonds forming between atoms of beyond-nearest-neighbor amino acids. Consequently, at this level, proteins can be viewed as a sequence of helices and beta strands (elements that form the pleated beta sheets) joined by less regular connections called loops and turns.

In natural conditions, the arrays of the secondary structures further fold to make space-filling configurations -- the native states of the proteins, which define their biological function. Predicting the shape of the native states from the sequence of amino acids is a very important, but formidable task. Due to the large number of degrees of freedom in this problem, current all-atom computer simulations only allow to fold relatively short chains, yet in time significantly longer than it takes a nature protein to do so. Different techniques are implemented to effectively reduce the number of degrees of freedom. Although the latter simulations are more efficient, they suffer from a large number of input parameters, which restrict their predictive power.

In Ref.~\onlinecite{Danielsson:2009qm,Chernodub:2010xz} it was proposed to apply an effective field theory approach to proteins, constructing effective energy functionals for 3D curves. It was argued that a natural description is in terms of a gauge theory that is a low-dimensional analog of the Abelian Higgs model. Such model would reproduce helices as ground state configurations and loops as solitons interpolating between the ground states. In a series of subsequent works\cite{Molkenthin:2011,Hu:2011,Hu:2011wg,Krokhotin:2012,Peng:2014,Niemi:2014,Molochkov:2017jmv} it was shown that a predominant number of secondary structures found in real proteins can be fit with a sub-angstrom accuracy by an effective model with an order of magnitude fewer number of parameters than that of the conventional coarse grained models.

In the papers using the effective theory approach the proteins were described by a discrete version of the model owing to the fact that proteins are discrete chains with no translational symmetry. In this letter we discuss the continuous Abelian Higgs model and explain a number of universal features of proteins that this model predicts. We will claim that the phenomenology is based on a two-minimum potential, whose ground state configurations are helices. Beta strands can appear as metastable configurations or as solitons, and their abundance is controlled by one of the parameters of the model. Moreover, we establish several relations between the geometry of different secondary structure elements. In the end we comment, how discreteness and inhomogeneity of proteins can be accounted in generalizations of the present model. The various details of the analysis summarized in this letter appeared in a companion paper.~\cite{Melnikov:2019}


Following Refs.~\onlinecite{Danielsson:2009qm,Chernodub:2010xz} we will use curvature $\kappa$ and torsion $\tau$ as the effective fields of the large-scale dynamics of the protein molecules. We will consider them as functions of the (arc) length parameter ($s$) of the curves. For these fields we will write an effective energy functional, cf.~Refs.~\onlinecite{Hu:2013,Gordeli:2015aya}.

A minimal model necessary for reproducing the features of the protein molecules, relevant for the present discussion, is the one-dimensional Abelian Higgs model with a Chern-Simons term and a Proca mass term:
\begin{multline}
\label{AHfunctional}
E  = \int\limits_0^L ds\ \frac{1}{2}\left(|\nabla\hk|^2 - m^2|\hk|^2 + \lambda|\hk|^4 \right)
\\ - F\int\limits_0^L ds\ \htau + \int\limits_0^L ds\ \frac{1}{2}\epsilon^2(\htau-\eta')^2\,.
\end{multline}
Here complex $\hk=\kappa e^{i\eta}$ and real $\htau$ denote fields, subject to local gauge transformations, so that physical gauge invariant curvature and torsion are $\kappa=|\hk|$ and $\tau=\htau-\eta'$. Quantities $m$, $\lambda$, $F$ and $\epsilon$ are phenomenological parameters, which can be fixed by comparing this model with real proteins. $L$ is the length of the curve, which will later be considered infinite.

We note that the first term in the second line of Eq.~(\ref{AHfunctional}), representing the one-dimensional Chern-Simons action, is not gauge invariant in the case of finite open curves. Consequently, the model possesses physical edge modes corresponding to the choice of normal vectors at the endpoints of the curves.~\cite{Melnikov:2019} 

The torsion field enters quadratically in the energy functional and can be integrated out assuming 
\be
\label{tkrel}
\tau \ = \ \frac{F^2}{\kappa^2+\epsilon^2}\,.
\ee
We also note that all the equations depend on gauge invariant quantities $\tau$ and $\kappa$. This situation is equivalent to the Higgs mechanism of the spontaneous breaking of the original $U(1)$ gauge theory. The reduced gauge invariant energy functional takes the following form,
\be
\label{EffAction}
E = \frac{1}{2}\int\limits_0^L ds \left({\kappa'}^2 - m^2\kappa^2 + \lambda\kappa^4 - \frac{F^2}{\kappa^2+\epsilon^2}\right),
\ee
with a residual $\mathbb{Z}_2$ symmetry $\kappa\to -\kappa$. 

The first feature of the model is special relation~(\ref{tkrel}) between the curvature and the torsion, which emphasizes the importance of the Chern-Simons term for the solutions with non-zero torsion. The second feature is a non-local effective potential for the curvature field obtained upon integration of the torsion. To highlight other features we discuss the classical minimum energy configurations of this theory. 

\begin{figure}[t]
\includegraphics[width=\linewidth]{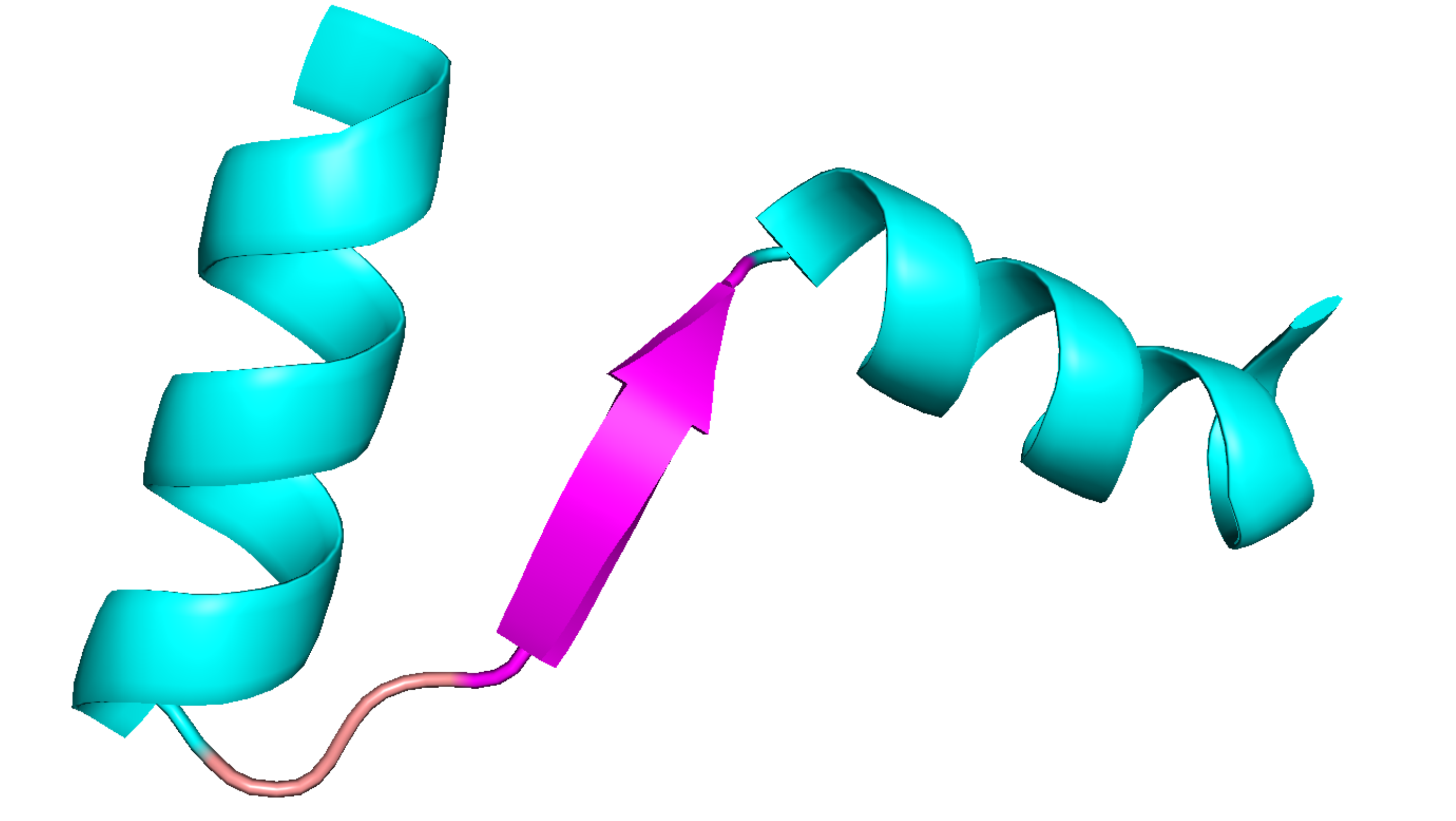}
 \caption{(Color online) Customary secondary structure representation of a piece of a protein chain. The cyan colored spirals are alpha helices and the twisted purple ribbon is a beta strand, here sandwiched between two helices. The image is produced with the help of the PyMOL software.\cite{pymol}}
 \label{fig:1g66}
\end{figure}

It is convenient to rewrite the derivative-free part of Eq.~(\ref{EffAction}) in terms of a potential,
\be
\label{potential}
V(\kappa) \ = \ \frac{\lambda(\kappa^2-\kappa_0^2)^2(\kappa^2+\kappa_1^2)}{2(\kappa^2+\epsilon^2)}\,,
\ee
with another set of phenomenological parameters $(\lambda,\kappa_0,\kappa_1,\epsilon)$. As explained in Ref.~\onlinecite{Melnikov:2019}, the most interesting scenario corresponds to $0\leq \kappa_1^2\leq \epsilon^2$. Let us review the properties of this potential in the limit of infinite curves. Finite size effects are important for the stability of possible configurations. They are discussed in some detail in Ref.~\onlinecite{Melnikov:2019}.

Static solutions in the infinite length model include a true ground state with constant $\kappa=\kappa_0$ and $\tau=\tau_0$, stable kink-like solutions, interpolating between the minima $\kappa_0$ and $-\kappa_0$, local energy minimum solutions $\kappa=0$ and unstable sphalerons of the local energy minimum. The last two solutions are only present if $\kappa_1^2\leq \kappa_0^2\epsilon^2/(\kappa_0^2+2\epsilon^2)$. The sphalerons~\cite{Klinkhamer:1984di,Rubakov} are classical bounce solutions of the particle motion in an inverted potential. They characterize the height of the potential barrier separating the true and the false ground states.

These solutions have a natural interpretation in terms of the observed secondary structures of proteins. The constant curvature, $\kappa=\kappa_0$, constant torsion curves are helices, which are the most abundant regular structures (for example, alpha helices). The configurations with zero curvature (but non-zero torsion) can be compared with beta strands in proteins -- zigzag-like configurations of the backbone chain, which are typically visualized by quasi straight and slightly twisted ribbons as in figure~\ref{fig:1g66}. The kinks interpolating between global minima are loops (structural motifs) connecting pairs of helices. Finally, sphalerons are non-zero curvature segments connecting two straight pieces. They are unstable in the model with translational symmetry, but can be stabilized by finite size and discretization effects. Thus they can be compared with higher curvature loops (hairpins) connecting the beta strands.


\begin{figure}[t]
\includegraphics[width=\linewidth]{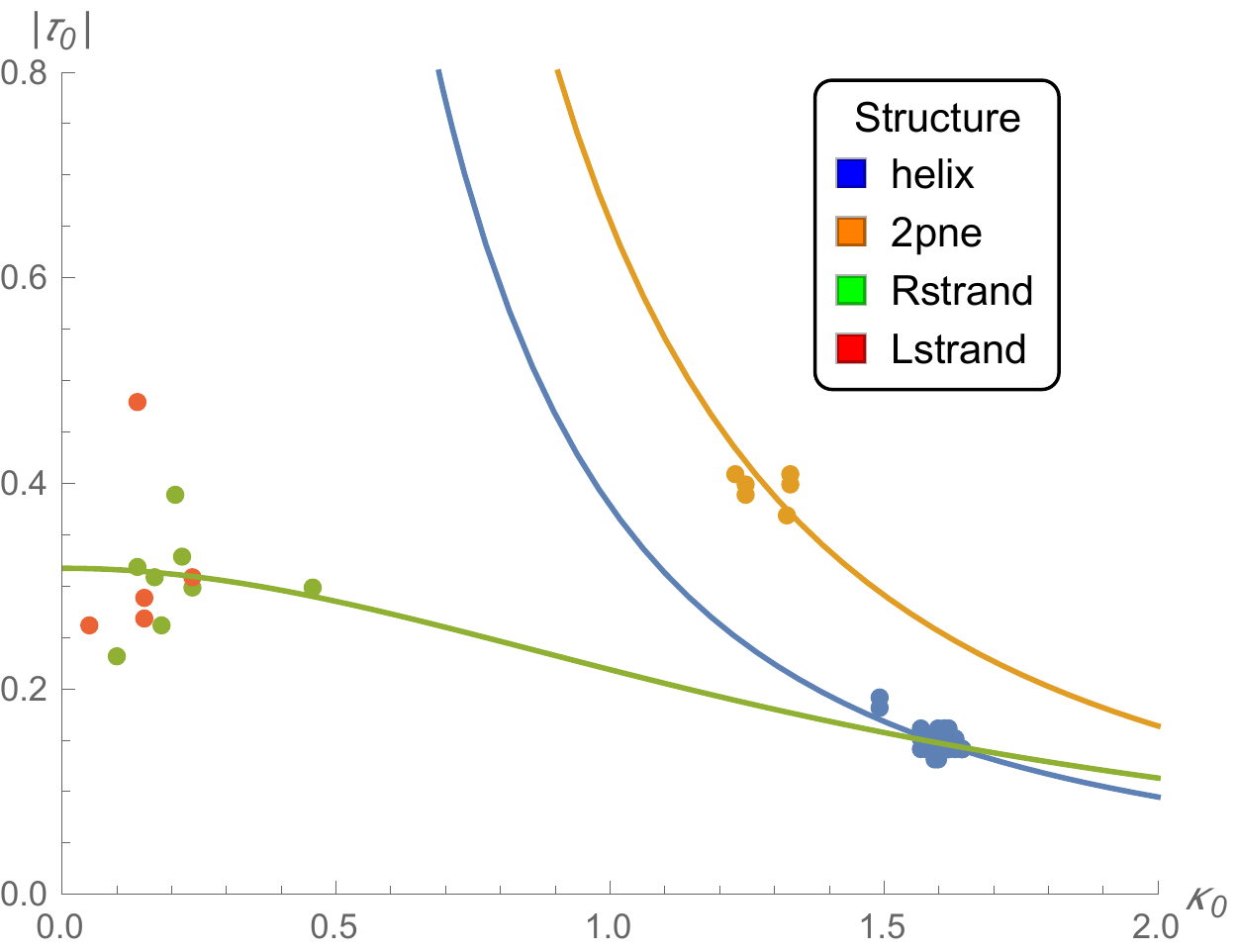}
 \caption{(Color online) Distribution of the curvature and torsion pairs. The color identification is blue for the alpha helices, green for the right-handed beta strands, red for the left-handed strands and orange for the strands of the \emph{2pne} protein. The green curve is a fit~(\ref{tkrel}) for the helix and strand points (excluding \emph{2pne}). The blue and orange curves are the same fits, but assuming $\epsilon=0$, for the helices and for the \emph{2pne}.}
\label{fig:kt}
\end{figure}

At the next level of the comparison of the model with the protein phenomenology, we can estimate the values of the parameters in the effective energy functional. In Ref.~\onlinecite{Melnikov:2019} we fitted the positions of the $C_\alpha$ atoms in the helices with continuous curves to obtain the following estimates for the curvature and torsion:
\be
\label{k0t0}
\kappa_0 \  \simeq \ 1.6~\text{\AA}^{-1}\,, \qquad \tau_0 \ \simeq \ 0.15~\text{\AA}^{-1}\,.
\ee

One can then ask whether relation~(\ref{tkrel}) is satisfied in real proteins. The relation roughly tells us that structures with lower curvature should have higher torsion. Alpha helices have a rather narrow distribution around values of Eq.~(\ref{k0t0}), so we checked whether beta strands can be fit in this picture. One can think of the ribbons, as the one representing the beta strand on figure~\ref{fig:1g66} as a stretched helix with $\kappa/\tau\ll 1$. By fitting the strands as such stretched helices we obtained a distribution shown on figure~\ref{fig:kt}.\cite{Melnikov:2019} We stress that in the case of the beta strands we measured the torsion of the ribbon, rather than that of the backbone chain.

We find that strands can have both positive and negative torsion as shown on figure~\ref{fig:kt}. Moreover there are some special strands that do not fit the present model, or rather the universality class of the specific relation~(\ref{tkrel}). We found such strands in the \emph{2pne} protein. The strand of that protein are bona fide left helices, but with non-standard curvature and torsion.

Figure~\ref{fig:kt} is similar in spirit to the famous Ramachandran plots\cite{Ramachandran:1963,Ramachandran:1965,Lovell:2003}, cast in the form, which allows to extract the information about relation~(\ref{tkrel}). As in the Ramachandran plots, the distribution of the beta strands is much less localized in comparison to alpha helices, but it is compatible with Eq.~\ref{tkrel}. In particular, it is consistent with the non-zero Proca mass, which acts an IR regulator in potential $(\ref{potential})$.  By fitting the data on figure~\ref{fig:kt} we can obtain the average values of parameters $F$ and $\epsilon$:
\be
F \ = \ 0.70~\text{\AA}^{-1}\,, \qquad  \epsilon \ = \ 1.5~\text{\AA}^{-1}\,. 
\ee

\begin{figure}[t]
\includegraphics[width=\linewidth]{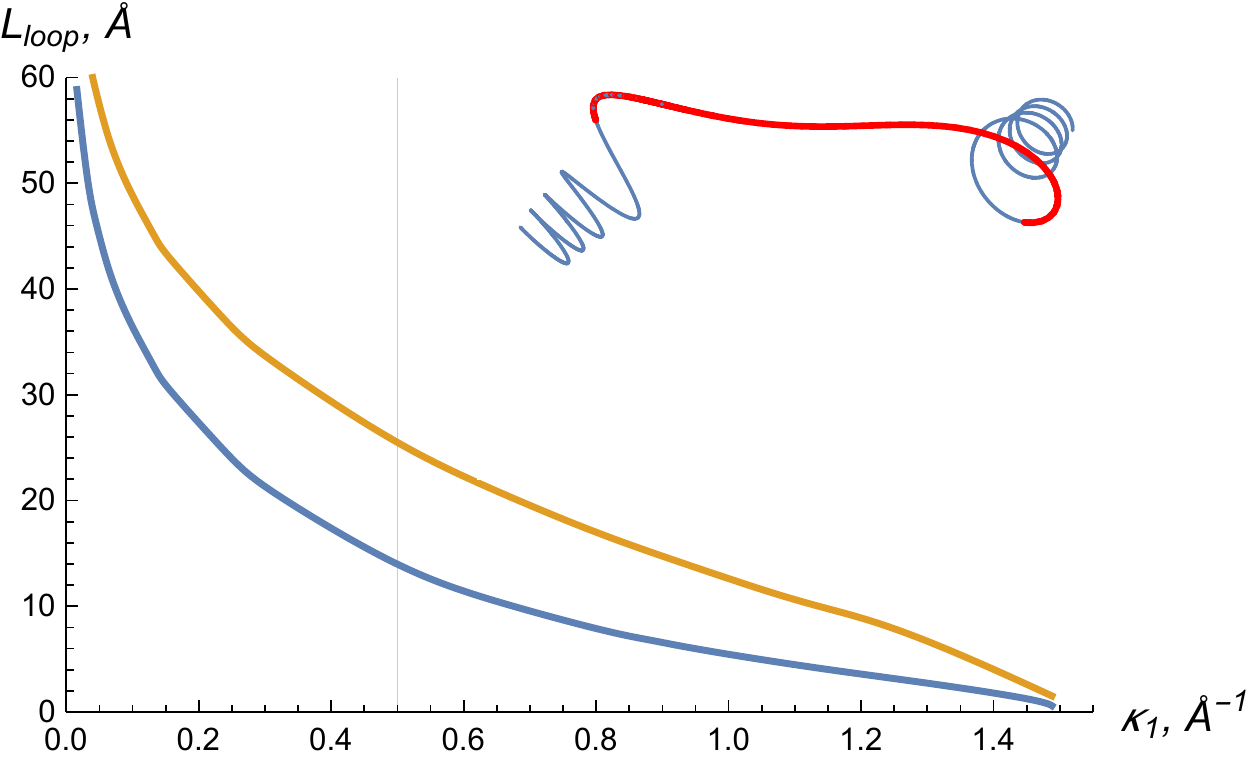}
 \caption{(Color online) Size of the kink interpolating between two global minima in potential~(\ref{potential}). The curves correspond to the sections containing 50\% (blue) and 90\% (orange) of the soliton energy. The inset shows the shape of the curve for $\kappa_1=0.5~\text{\AA}^{-1}$. Part of the curve highlighted in red, shows the piece contributing 50\% of the energy. }
\label{fig:loopsize}
\end{figure}

\begin{figure}[t]
\includegraphics[width=\linewidth]{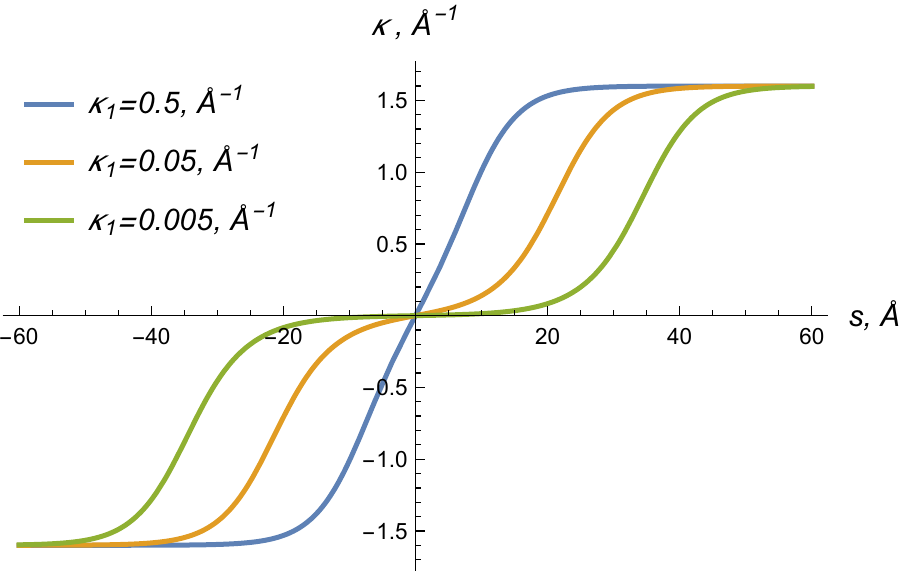}
 \caption{(Color online) Kink solutions of the model with potential~(\ref{potential}). For small $\kappa_1\to 0$ the size of the step in the middle grows logarthmically.}
\label{fig:kinks}
\end{figure}

With $\kappa_0$, $F$ and $\epsilon$ fixed there remains only one free parameter in the model. We can choose it to be $\lambda$ or $\kappa_1$. The remaining parameter controls the size of the solitons. In figure~\ref{fig:loopsize} we show how the size of the kink depends on $\kappa_1$. In general, we see a logarithmic decrease of the loop size with increase of $\kappa_1$. For small $\kappa_1$ one observes long loops, which correspond to a characteristic step appearing in the kink solution, as can be seen on figure~\ref{fig:kinks}.\footnote{As far as we are aware, similar kink solutions were considered in Refs.~\onlinecite{Christ,Bazeia,Bazeia:2019vld}.} In the limit $\kappa_1\to 0$, the global and the local minima become degenerate and the original kink splits into a pair of kinks with infinite separation.

The step appearing in the kink solution in figure~\ref{fig:kinks} is a piece of the curve with much lower curvature (see also the inset on figure~\ref{fig:loopsize}), which should be interpreted in terms of the configuration shown on figure~\ref{fig:1g66}: a pair of alpha helices separated by a beta strand. One can make the following estimate of the size of the small curvature region, assuming that it is characterized by $|\kappa| \ll \kappa_1\ll \kappa_0$,
\be
R_{\beta} \simeq  \int \limits_{-\kappa_1}^{\kappa_1} \frac{dk}{\sqrt{\lambda}\kappa_0^2}\frac{\epsilon}{\sqrt{k^2+\kappa_1^2}} \simeq  \frac{2\epsilon^2(\kappa_0^2+\epsilon^2)}{F\kappa_0^2}\,,
\ee
where we assumed the relation between parameters $\lambda$ and $F$ following from two different parameterizations of the potential.\cite{Melnikov:2019} Note, that this value does not depend on $\kappa_1$. The estimate gives the numerical value $R_\beta\simeq 12$~\AA, which is a prediction of the universal size of the length of the beta strand in the model. Beta strands can appear longer, because they have an attached intermediate region, which depends logarithmically on $\kappa_1$.

On the other hand, if $\kappa_1$ is not much smaller than $\kappa_0$, the step is not formed and the configurations like the one on figure~\ref{fig:1g66} are not possible. Hence $\kappa_1$ can be viewed as an external parameter, like chemical potential, controlling the ability of the protein to form beta strands. This chemical potential can either be a characteristic of the medium, in which the protein is present, or of the amino acid composition of the backbone chain.

It is interesting to discuss what happens in the limit $\kappa_1\to 0$ in more detail. In this limit the minima of the potential are almost degenerate and the kinks are characterized by long, almost straight, segments inserted between the helices. In Ref.~\onlinecite{Melnikov:2019} we also made an estimate of the stability of the metastable configuration with $\kappa=0$. Such structures become more stable in the $\kappa_1\to 0$ limit. Since actual proteins are finite and discrete, it is also plausible that the steps of the kinks would be ``mixing" up with $\kappa=0$ segments. In other words, it is plausible that  longer chains of the beta strands are formed in that regime. The sphalerons might also stabilize in the discrete case. They would introduce loops of higher curvature connecting beta strands (similar to the known hairpin motifs in proteins). 

There is a way how the continuous model can be deformed to account for the breaking of the translational invariance in actual proteins. Apart from considering finite curves, one can introduce coordinate dependence to the parameters. Apart from a discrete periodicity of the backbone chain, this could also take into account the local inhomogeneity of the chemical structure. It is then natural to fit the continuous curves of real proteins using a Fourier expansion. We hope to discuss such generalizations in a future work.

\paragraph*{Acknowledgements}  DM would like to thank Antti Niemi and Ara Sedrakyan for many useful discussions on the application of effective field theory and topology to protein physics. DM is also grateful to Dionisio Bazeia, Sergei Brazovskii and the participants of the workshop ``Physics and Biology of Proteins'' held at the International Institute of Physics in Natal in June 2017 for interesting ideas and discussions. This work was supported by the grant No.~16-12-10344 of the Russian Science Foundation.

\end{document}